\documentclass[sigconf]{acmart}

\AtBeginDocument{%
  }

\setcopyright{acmlicensed}
\copyrightyear{2024}
\acmYear{2024}
\acmDOI{XXXXXXX.XXXXXXX}

\acmConference[Conference acronym 'XX]{Make sure to enter the correct
  conference title from your rights confirmation emai}{2024}{XXX}
\acmISBN{XXXXXXXX}

\usepackage[utf8]{inputenc} 
\usepackage[T1]{fontenc}    
\usepackage{hyperref}       
\usepackage{url}            
\usepackage{booktabs}       
\usepackage{amsfonts}       
\usepackage{amsmath}
\usepackage{nicefrac}       
\usepackage{microtype}      
\usepackage{xcolor}         
\usepackage{multirow}
\usepackage{float}
\usepackage{graphicx}
\usepackage{multicol}
\newcommand{\N}{\mathbb{N}}
\newcommand{\R}{\mathbb{R}}
\newcommand{\seq}{\mathbf{u}}
\newcommand{\len}{\ell}
\newcommand{\dime}{{d_e}}
\newcommand{\dimr}{{d_r}}
\newcommand{\dimp}{{d_p}}
\newcommand{\bs}{b}
\newcommand{\bU}{\mathbf{U}}
\newcommand{\lifted}{\mathbf{Y}}
\newcommand{\BT}{\text{BT}}
\newcommand{\bl}{\mathbf{y}}
\newcommand{\corr}{\mathcal{C}}
\newcommand{\eqdef}{:=}

\begin{document}

\title{Enhancing User Sequence Modeling through\\Barlow Twins-based Self-Supervised Learning}

\author{Yuhan Liu}
\authornote{Work was completed during an internship at Google Research}
\email{yl2976@cornell.edu}
\affiliation{
  \institution{Cornell University}
  \country{USA}
}

\author{Lin Ning}
\authornote{Corresponding author.}
\email{linning@google.com}
\affiliation{
  \institution{Google Research}
  \country{USA}
}

\author{Neo Wu}
\email{neowu@google.com}
\affiliation{
  \institution{Google Research}
  \country{USA}
}

\author{Karan Singhal}
\email{karansinghal@google.com}
\affiliation{
  \institution{Google Research}
  \country{USA}
}

\author{Philip Andrew Mansfield}
\email{memes@google.com}
\affiliation{
  \institution{Google Research}
  \country{Canada}
}

\author{Devora Berlowitz}
\email{devorab@google.com}
\affiliation{
  \institution{Google}
  \country{USA}
}

\author{Sushant Prakash}
\email{sush@google.com}
\affiliation{
  \institution{Google DeepMind}
  \country{USA}
}

\author{Bradley Green}
\email{brg@google.com}
\affiliation{
  \institution{Google DeepMind}
  \country{USA}
}


\begin{abstract}
User sequence modeling is crucial for modern large-scale recommendation systems, as it enables the extraction of informative representations of users and items from their historical interactions. These user representations are widely used for a variety of downstream tasks to enhance users' online experience. A key challenge for learning these representations is the lack of labeled training data. While self-supervised learning (SSL) methods have emerged as a promising solution for learning representations from unlabeled data, many existing approaches rely on extensive negative sampling, which can be computationally expensive and may not always be feasible in real-world scenario. In this work, we propose an adaptation of Barlow Twins, a state-of-the-art SSL methods, to user sequence modeling by incorporating suitable augmentation methods. Our approach aims to mitigate the need for large negative sample batches, enabling effective representation learning with smaller batch sizes and limited labeled data. We evaluate our method on the MovieLens-1M, MovieLens-20M, and Yelp datasets, demonstrating that our method consistently outperforms the widely-used dual encoder model across three downstream tasks, achieving an 8\%-20\% improvement in accuracy. Our findings underscore the effectiveness of our approach in extracting valuable sequence-level information for user modeling, particularly in scenarios where labeled data is scarce and negative examples are limited.
\end{abstract}



\maketitle

\section{Introduction}
\label{sec:introduction}
Embedding-based deep neural networks (DNNs) has become pivotal in large-scale recommendation systems \citep{covington2016deep, Ma2018KDD, Yi2019RecSys, Yang2020WWW, Jiang2020WWW, Volkovs2017NIPS,DeepLearningRecSysSurvey2019}, learning user and item representations from vast amounts of user-item interaction data to power various downstream tasks like predicting user preferences, learning user demographics, and recommending relevant items. However, the scarcity of high-quality labeled user data poses a significant challenge for supervised learning approaches.

Labeled user data, such as demographics, interests, or specific item preferences, is crucial for training accurate predictive models. However, despite the abundance of user interaction data, obtaining high-quality labels is difficult due to several factors. Privacy concerns often restrict the use of sensitive user data, even with paid human annotators. Furthermore, user preferences are inherently subjective and difficult to label consistently, as individual interpretations vary. Explicitly soliciting feedback through surveys or ratings often yields low response rates and biased results, further exacerbating the scarcity of reliable labels. These challenges hinders the development of sophisticated personalization models, particularly for new users with limited interaction histories or in rapidly changing environments where user preferences evolve quickly.

Self-supervised learning (SSL) offers a promising solution by learning informative representations from large unlabeled data through pretext tasks and data transformations. It encourages the model to learn meaningful patterns within the data itself and create representations that capture the essential underlying information invariant to the data transformations. SSL has achieved notable success in various domains such as computer vision~\cite{ChenH21,ChenK0H20,ZbontarJMLD21,GrillSATRBDPGAP20} and natural language processing (NLP)~\cite{DevlinCLT19,liu2019roberta,BrownMRSKDNSSAA20,openai2023gpt4,baevski23a}, and it is a major driving force in recent advances of powerful foundation models~\cite{bommasani2021opportunities,openai2023gpt4}. While existing research \cite{yu2023selfsupervised} has explored applying SSL to recommendation systems and user modeling , challenges remain in adapting these methods effectively due to the unique characteristics of user sequence data.

A key challenge in adapting SSL for user sequences is the reliance of many existing methods on extensive negative sampling. Contrastive learning, for instance, while effective in vision~\cite{ChenH21} and NLP tasks~\cite{GaoYC21,MengXBTBHS21,neelakatan2022text}, often requires large batches with numerous negative examples when applying to recommendation systems~\cite{Yao2021self,xie2022contrastive,ChenLLMX22,XiaHHLYK23}, increasing computational costs and posing challenges in real-world scenarios with limited negative samples. Dual encoders, commonly used in recommendation systems for learning sequence-level representations, also require large amount of negative examples and are often task-specific and may not generalize well.


In this work, we focus on adapting Barlow Twins~\cite{ZbontarJMLD21}, a state-of-the-art SSL method, to user sequence modeling. Barlow Twins is particularly appealing due to its ability to learn effective and generalized representations without relying on negative sampling, which addresses the key limitations of many existing SSL methods. While Barlow Twins has primarily been applied to highly redundant data like images and audio, we demonstrate that, with suitable augmentation methods tailored for user sequences, it can effectively learn meaningful sequence-level representations for a variety of downstream tasks, even in the absence of labeled data or with limited negative samples.

\textbf{Key contributions:}
\begin{enumerate}
    \item We demonstrate the first successful adaptation of Barlow Twins to low-redundant user sequence data, a domain significantly different from its typical applications in image~\cite{ZbontarJMLD21} and audio processing~\cite{anton2023audio}. This adaptation unlocks the potential for learning informative representations of user behavior without relying on labeled data or extensive negative examples.
    \item We show that our Barlow Twins-based representations consistently outperform those learned by traditional dual-encoder models trained for next-item prediction, particularly in scenarios with limited labeled data, highlighting the effectiveness and generalizability of our approach.
    \item Our approach offers distinct advantages over prevalent SSL methods for user sequence modeling. It eliminates the need for computationally expensive negative sampling, demonstrates robustness to small batch sizes, and naturally avoids trivial (constant) embeddings \cite{ZbontarJMLD21}.
    \item We provide a thorough quantitative analysis of our approach across a range of downstream tasks, including next-item prediction and sequence-level classification. Additionally, we systematically examine the impact of various data augmentation methods on the performance of our SSL framework.
\end{enumerate}

\section{Related work}

\begin{figure*}
    \centering
    \includegraphics[width=0.6\linewidth]{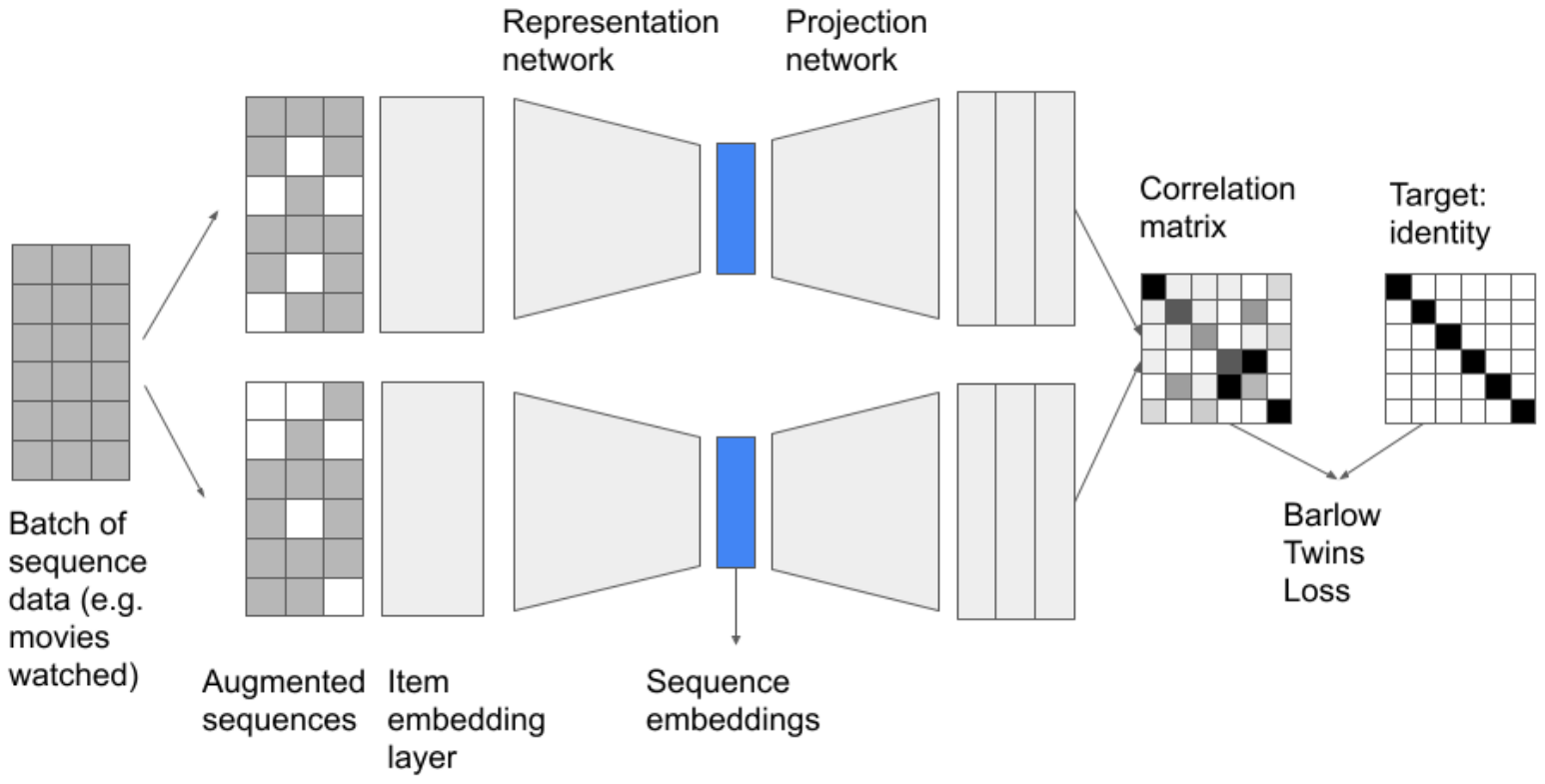}
    \caption{Illustration of Barlow Twins for user sequence modeling. Two independent augmentations are applied to the same batch, and the loss function enforces statistically independent components.}
    \label{fig:barlow-twins}
\end{figure*}

\subsection{Self-supervised Learning}
Self-supervised representation learning approaches can be broadly categorized as generative and discrimative~\cite{ChenK0H20,GrillSATRBDPGAP20}. Generative approaches include adversarial training~\cite{GoodfellowPMXWOCB14} and reconstruction-based approaches~\cite{rezende14,PathakKDDE16,HeCXLDG22}. The latter is very effective in large vision and language models. In computer vision, \cite{HeCXLDG22} learns representations by reconstructing masked images. For NLP tasks, large language models like BERT~\cite{DevlinCLT19,liu2019roberta} and GPT~\cite{BrownMRSKDNSSAA20,openai2023gpt4} often use masked language modeling which predicts masked tokens in the input sequence. 
These models usually have enormous parameters and require significant computational resources to train from scratch. There are several works that enable learning representations with smaller batch sizes and less computation~\cite{MengXBTBHS21,LiuVKC21}, but require pretrained weights from large language models. 

Discriminative approaches, on the other hand, avoid the computationally expensive generation process. These methods can involve designing input-specific prediction tasks like coloring gray-scale images~\cite{LarssonMS16} or predicting the relative patch positions, and motion prediction~\cite{DoerschZ17} in vision. In large language models~\cite{DevlinCLT19,liu2019roberta}, next sentence prediction is used in conjunction with masked language modeling to learn semantic relationships between sentences.

Contrastive learning, popular in both vision~\cite{ChenK0H20} and NLP tasks~\cite{GaoYC21,MengXBTBHS21,neelakatan2022text}, learns representations by bringing similar examples closer and dissimilar examples further apart. However, this approach usually requires large training batches with many negative samples, which can be computationally challenging and infeasible when negative examples are limited. Alternative approaches like Siamese networks or the Barlow Twins loss have been proposed to mitigate the reliance on negative samples. \cite{GrillSATRBDPGAP20,ChenH21} use Siamese networks~\cite{BromleyBBGLMSS93} on two views of the same image and apply special operations (momentum encoder, stop gradient) on one branch to prevent trivial representations. \cite{ZbontarJMLD21} uses a Barlow Twins loss to learn representations with statistically independent components.

For sequential data beyond natural language, \cite{BaiWZX21} applied SSL to continuous time series data with the predictive information objective, which captures the mutual information between past and future events. The objective is hard to compute exactly, and the authors had to rely on stationarity and Gaussian assumptions, which are unlikely to hold for sequences over large discrete domains. While Barlow Twins has been successfully applied to audio inputs \cite{anton2023audio}, the spectral domain augmentations used in that work are not directly applicable to user sequence data. Therefore, our adaptation of Barlow Twins to user sequence modeling represents a significant technical contribution.

\subsection{SSL for User Sequence Modeling}
Techniques from self-supervised learning have been successfully applied to recommendation systems. For sequential recommendation, \cite{SunLWPLOJ19} proposed BERT4Rec, which adopts BERT~\cite{DevlinCLT19} model for recommendation systems. \cite{ZhouWZZWZWW20} designed auxiliary self-supervised tasks that learn correlations among attributes, items, and sequences. A series of works~\cite{Yao2021self,xie2022contrastive,ChenLLMX22,XiaHHLYK23} applied contrastive learning to improve recommendation performance. However, their evaluation was solely on next item prediction and did not consider other tasks that could potentially benefit from the user representations. Several works~\cite{Cai0XR23,yang2023debiased} applied graph representation learning to user-item and user-user interaction graphs using graph neural networks. This line of work is orthogonal to ours as these works leverage other information from interaction graphs. We believe a better user sequence representation can potentially improve these graph-based methods.

\section{Method}
\subsection{User Sequence Model}
We assume that users can perform an action from a finite discrete domain $\mathcal{D}$. A user sequence model $U:\mathcal{D}^\len\mapsto\R^{\dimr}$ takes a sequence of user actions $$\seq=(u_1, \ldots, u_\ell)\in\mathcal{D}^\len$$ with length $\len$ as an input, where each $u_i$ is a unique integer identifier that uniquely representing an action (e.g. a movie watched by the user). The output is a $\dimr$-dimensional vector representation of the sequence.

To obtain this representation, $\seq$ is first passed through an item embedding layer $E:\N^\ell\mapsto \R^{\dime\times \len}$, transforming each integer ID into a $d_e$-dimensional embedding vector. A representation network $R:\R^{\dime\times \ell}\mapsto\R^{\dimr}$ then processes the sequence of embeddings to produce the final sequence-level representation with $\dimr$ dimensions. Thus, the user sequence model can be expressed as $$U=R\circ E.$$ 

While the choice of representation network $R$ is flexible, we use a simple convolutional neural network (CNN) for simplicity as we primarily focus on demonstrating the effectiveness of Barlow Twins-based SSL on downstream tasks. In practice, more sophisticated architectures like Transformers~\cite{DBLP:conf/nips/VaswaniSPUJGKP17} could be employed for potentially better performance. For all downstream tasks, the user sequence model $U$ serves as a base model to process the input sequence and output a sequence level representation, which are then passed to task-specific neural networks.

\subsection{Barlow Twins for User Sequence Data}

Figure~\ref{fig:barlow-twins} illustrates our adaptation of Barlow Twins to user sequence modeling. The model consists of two branches with shared weights that process two views of the same input batch, with a final Barlow Twins loss applied to the outputs of two branches.

Specifically, each branch comprises a sequence representation network $U$ and a projection network 
$P:\R^{\dimr}\mapsto\R^{\dimp}$,
which is an MLP with $\dimp>\dimr$ that maps the sequence-level representation obtained from $U$ to a higher dimensional space. We denote the model with projection layers as 
$$\BT := P\circ R\circ E.$$

During self-supervised pretraining, for each batch of sequences $\bU = [\seq_1, \ldots, \seq_{\bs}]$, two independent augmentations are applied, yielding two augmented batches $\bU_1, \bU_2$ (augmentation methods are detailed in Section~\ref{sec:augmentation}). These augmented batches are then passed through the two branches of $\BT$ with shared weights. The resulting outputs, denoted by $\lifted^i =[\bl^i_{1}, \ldots, \bl^i_{\dimp}], i=1,2$, are mean-centered along the batch dimension .

We minimize the Barlow Twins loss,
\begin{equation}
    \mathcal{L}_{BT}\eqdef \sum_{i=1}^{\dimp}(1-\corr_{ii})+\lambda \sum_{i\ne j}\corr_{ij}^2,
\end{equation}
where
\begin{equation}
    \corr_{ij}\eqdef \frac{\sum_{j=1}^{\bs}\bl^1_{1,j}\bl^1_{2,j}}{\sqrt{\sum_{j=1}^{\bs}(\bl_{1,j})^2}\sqrt{\sum_{j=1}^{\bs}(\bl_{2,j})^2}}
\end{equation}
is the cross correlation matrix along the batch dimension, and $\lambda$ is a hyperparameter balancing the two terms. This loss enforces $\corr$ to be close to an identical matrix, guiding the model to learn statistically independent components in the representation. 

\subsection{Augmentation Methods}
\label{sec:augmentation}
In our experiments, we investigate the impact of three different data augmentation techniques on the performance of our Barlow Twins-based user sequence model:
\begin{enumerate}
    \item {\bf Random masking (RM)}: Each item in the sequence is independently replaced with a mask token $[\text{mask}]$ with probability $p\in(0, 1)$. This technique encourages the model to learn to infer missing information from the surrounding context.
    \item {\bf Segment masking (SM)}: A contiguous subsequence of length $\lfloor p\ell\rfloor$, where $p\in(0, 1)$, is randomly selected and all items within that subsequence are replaced with the mask token $[\text{mask}]$. This encourages the model to learn longer-range dependencies and contextual information.
    \item {\bf Permutation}: The order of items in the input sequence is randomly permuted. This augmentation is particularly relevant for downstream tasks where the absolute position of an item in the sequence is less important than the overall composition of items.
\end{enumerate}

\subsection{Downstream Tasks}
\label{sec:downstream}
After pretraining, we discard the projection network and retain only the sequence representation model, $U$. For each downstream task, We then append task-specific network structure on top of $U$.

\paragraph{{\bf Sequence-level classification}} For sequence-level classification tasks, we add a 2-layered multi-layer perceptron (MLP) head to the base model $U$. The output dimension of this MLP is set to match the number of categories in the specific classification task.

\paragraph{{\bf Next item prediction}} For next-item prediction, a canonical task in sequential recommendation, we construct a dual-encoder model on top of $U$. The context tower consists of $U$ augmented with a two-layer MLP, which projects the sequence-level representation into the item embedding space. The item tower is simply the item embedding $E$. The model is trained using a contrastive loss function.

During downstream task training, the weights of the representation model $U$ can be either fixed or trainable. Fixing the weights allows us to directly assess the quality of the pre-trained representations. On the other hand, making the weights trainable enables us to potentially achieve optimal performance when abundant labeled data is available for the specific downstream task.
\section{Experiment Setup}
We evaluate our Barlow Twins-based SSL pipeline on three datasets: Movielens 1M, Movielens 20M~\cite{HarperK16}, and Yelp. For each dataset, we first pre-train a user sequence model using our Barlow Twins adaptation, and then evaluate its effectiveness on various downstream tasks. We compare the performance of our Barlow Twins model to a dual encoder baseline model trained exclusively for next-item prediction.

\subsection{Datasets} The Movielens 1M dataset contains approximately 1 million movie ratings from 6040 users across 3952 movies. Each movie is associated with a unique movie ID, title, year, and a list of genres. Each user is labeled with gender, age group, and occupation. The Movielens 20M dataset is a larger version, containing 20 million movie ratings across 27278 movies from 138493 users. The Yelp dataset contains user check-in and reviews for 150346 businesses. We use a small subset which contains 908915 tips made by 301758 users.

For training and evaluation, we segment each user's interaction history into sequences of length 16, filtering out users with fewer than 10 actions. Sequences shorter than 16 are padded with a $[\text{mask}]$ token. Note that only item IDs are used in the sequences, discarding additional item attributes. The train-validation-test split is 80\%-10\%-10\%. The processed dataset sizes are summarized in Table~\ref{tab:dataset}.

\begin{table}
    \centering
    \begin{tabular}{|c|c|c|c|}
    \hline
         &  MovieLens-1M & MovieLens-20M & Yelp tips\\ \hline
         \# Users & 6040 & 138493 & 301758\\
         \# Items &  3952 & 27278  & 150436\\
         \# Actions& $\sim10^6$ & $\sim2\times 10^7$ & 908915\\
         \# Categories & 18 & 18 & 1311\\ \hline
        Train & 795335 & 10776260 & 253317\\
        Val and test & 99417 & 1347032 & 28146 \\ \hline
    \end{tabular}
    \caption{Dataset statistics and the number of train/val/test sequences after preprocessing.}
    \label{tab:dataset}
\end{table}

\subsection{Pre-training} 
We pre-train a user sequence model on the training set using both our Barlow Twins adaptation and a dual encoder baseline. The item embedding (i.e., embedding for the movies) dimension is set to 16, and we vary the batch size across 128, 256, 512, and 1024 to assess its impact on performance.

{\bf Barlow Twins model} It utilizes a 2-layer 1D-CNN as the sequence representation network $U$, with each layer consisting of 32 convolution filters of size 3 followed by max pooling of size 3. The projection network is a 2-layer MLP with hidden dimension of is $[256, 256]$. We set the trade-off parameter $\lambda$ to 10. For augmentation, we explore random masking with probabilities $p\in \{0.2,0.4,0.6,0.8\}$, segment masking with $p=0.2$, and permutation.

{\bf Dual encoder baseline} Itcomprises a context tower and an item tower. The context tower uses the same item embedding and representation network structure as the Barlow Twins model, followed by a 2-layer MLP with hidden dimensions of 32 and 16, respectively. The item tower is simply the item embedding layer, sharing weights with the context tower. During training, a batch of user sequences is fed to the context tower, and the corresponding ground truth next items are fed to the item tower. We then compute and minimize the contrastive loss between the outputs of the two towers during training.

\subsection{Downstream Evaluation}
\subsubsection{Tasks}
We evaluate the quality of the learned sequence representations on two types of tasks: 
\begin{itemize}
    \item {\bf Sequence-level classification}: We assess the model's ability to predict sequence-level properties, using prediction accuracy as the metric. It includes:
    \begin{itemize}
        \item \textbf{Favorite category prediction}: Predict the most frequent movie genre (MovieLens 1M/20M) or business category (Yelp) in the user's interaction sequence. MovieLens datasets have 18 genres, while Yelp has 1000+ categories.
        \item \textbf{User classification (MovieLens-1M only)}: Predict a user's age group (7 categories) and occupation (21 categories) based on the interaction sequence.
    \end{itemize}
    \item {\bf Next-item prediction}: We evaluate the model's ability to recommend the next item given a sequence of user's history interactions, using top-$k$ recall (or hit-ratio) with $k=1, 5, 10$ as the metric.
\end{itemize}

Due to the limited availability of user attributes, we perform only favorite genre/category and next item prediction for MovieLens 20M and Yelp.

\subsubsection{Model Architecture and Training Setup}
For sequence classification tasks, we utilize the pre-trained Barlow Twins and dual encoder models up to the sequence embedding layer. We then add a 2-layer MLP with 20 hidden units and an output layer matching the number of label categories for the specific task. The downstream task models are then finetuned, with the sequence representation layers ($U$) either fixed (to evaluate representation quality) or trainable (to potentially achieve optimal performance with abundant data). As a baseline, we train an equivalent model with the same architecture (i.e., item embedding, 2-layer CNN, 2-layer MLP) from scratch on each downstream task.

To assess performance under limited labeled data scenarios, we finetune the sequence-level classification models using only a small proportion of training data: 1\% for MovieLens-1M, 1\%,0.1\%,0.01\% for MovieLens-20M, and 5\%,1\% for Yelp tips. Validation is always performed on the \textit{full} validation set. Note that with 1\% of the training data, the amount of data used for training is significantly less than the validation/test dataset. We use a batch size of 64 for all experiments.

For the next-item prediction task, we initialize a dual encoder model with the pre-trained Barlow Twins weights for both the item embedding and sequence representation layers. This model is then finetuned on the full training set, with either fixed or trainable sequence embedding layers ($U$). The performance of this finetuned model is compared to the original dual encoder baseline trained from scratch.

\section{Results}
\subsection{Sequence-level Classification}

\begin{table*}
    \centering
    \begin{tabular}{|c | c | c | c |c | c | c | c | c | c | c |}
    \hline
         Task  & SSL BS & RM Train & RM Fixed & SM Train & SM Fixed & Per Train & Per Fixed & DE train & DE fixed & Baseline\\ \hline
         \multirow{4}{*}{FG } & 128 & 0.8247 & 0.8405 & 0.8474 & \textbf{0.861} & 0.7984 & 0.8002 & 0.7325 & 0.7133 & \multirow{4}{*}{0.7350} \\
         & 256 & 0.8235 & 0.8462 & 0.8392 & \textbf{0.8511} & 0.7968 & 0.8003 & 0.7460 & 0.7077 &\\
         & 512 & 0.8100 & 0.8402 & 0.8375 & \textbf{0.8465} & 0.7954 & 0.7949 & 0.7549 & 0.7072 &\\
         & 1024 & 0.8222 & \textbf{0.8505} & 0.8485 & 0.8405 & 0.7844 & 0.7812 & 0.7405 & 0.7049 &\\\hline
        \multirow{4}{*}{Occ} & 128 & 0.1534 & 0.154 & 0.1471 & \textbf{0.1548} & 0.1359 & 0.1523 & 0.1384 & 0.1355 & \multirow{4}{*}{0.1407} \\
         & 256 & 0.1430 & \textbf{0.1558} & 0.1403 & \textbf{0.1558} & 0.1483 & 0.1557 & 0.1324 & 0.1361 &\\
         & 512 & \textbf{0.1563} & 0.1558 & 0.1446 & 0.1535 & 0.1488 & 0.1533 & 0.1345 & 0.1324 &\\
         & 1024 & 0.1517 & 0.1548 & 0.1508 & 0.154 & 0.1457 & \textbf{0.1556} & 0.1402 & 0.133 &\\\hline
         \multirow{4}{*}{Age} & 128 & \textbf{0.4055} & 0.4015 & 0.4035 & 0.4004 & 0.4 & 0.4001 & 0.4017 & 0.397 & \multirow{4}{*}{0.3992} \\
         & 256 & \textbf{0.4078} & 0.3998 & 0.4 & 0.4002 & 0.4023 & 0.4001 & 0.4034 & 0.397 &\\
         & 512 & \textbf{0.4112} & 0.4017 & 0.4092 & 0.4004 & 0.4016 & 0.3975 & 0.4011 & 0.397 &\\
         & 1024 & 0.403 & 0.3993 & \textbf{0.4079} & 0.4002 & 0.3996 & 0.3992 & 0.3973 & 0.397 &\\\hline
    \end{tabular}
    \caption{Best validation accuracy of different models after training on 1\% training data on MovieLens-1M. FG: favorite genre, Occ: occupation, SSL BS: SSL batch size, RM: random masking, SM: segment masking, Per: permutation, DE: dual encoder, Train: trainable. The highest accuracy in each row is in bold.}
    \label{tab:results}
\end{table*}

\begin{table*}
    \centering
    \begin{tabular}{|c | c | c | c |c | c | c | c | c | c |}
    \hline
         Task  & SSL BS & R0.2 Train & R0.2 Fixed & R0.4 Train & R0.4 Fixed & R0.6 Train & R0.6 Fixed & R0.8 Train & R0.8 Fixed \\ \hline
         \multirow{4}{*}{FG } & 128 & 0.8247 & \textbf{0.8405} & 0.8047 & 0.8135 & 0.799 & 0.7366 & 0.7653 & 0.7411 \\
         & 256 & 0.8235 & \textbf{0.8462} & 0.8163 & 0.7989 & 0.7766 & 0.7604 & 0.7612 & 0.6694 \\
         & 512 & 0.8100 & \textbf{0.8402} & 0.8062 & 0.7953 & 0.7742 & 0.7328 & 0.7503 & 0.6747 \\
         & 1024 & 0.8222 & \textbf{0.8505} & 0.7994 & 0.7928 & 0.7735 & 0.7236 & 0.7768 & 0.6591 \\\hline
        \multirow{4}{*}{Occ} & 128 & 0.1534 & \textbf{0.154} & 0.1389 & 0.1523 & 0.145 & 0.1515 & 0.1464 & 0.1411 \\
         & 256 & 0.1430 & \textbf{0.1558} & 0.1462 & 0.1505 & 0.1406 & 0.1491 & 0.1348 & 0.1434 \\
         & 512 & \textbf{0.1563} & 0.1558 & 0.1391 & 0.1516 & 0.134 & 0.15 & 0.1426 & 0.1408 \\
         & 1024 & 0.1517 & \textbf{0.1548} & 0.1522 & 0.1544 & 0.1402 & 0.1481 & 0.138 & 0.1401 \\\hline
         \multirow{4}{*}{Age} & 128 & \textbf{0.4055} & 0.4015 & 0.4026 & 0.3974 & 0.4001 & 0.3999 & 0.4089 & 0.3978 \\
         & 256 & \textbf{0.4078} & 0.3998 & 0.402 & 0.3989 & 0.403 & 0.3979 & 0.3975 & 0.3973 \\
         & 512 & \textbf{0.4112} & 0.4017 & 0.4086 & 0.3978 & 0.4106 & 0.3989 & 0.4083 & 0.3973 \\
         & 1024 & 0.403 & 0.3993 & \textbf{0.413} & 0.3995 & 0.4078 & 0.3977 & 0.4026 & 0.3973 \\\hline
    \end{tabular}
    \caption{Best validation accuracy of random masking with different masking ratios after training on 1\% training data on MovieLens 1M. R0.2, R0.4, R0.6, and R0.8 refer to the masking ratios of 0.2, 0.4, 0.6, and 0.8, respectively.}
    \label{tab:results-RM-ratios}
\end{table*}

\begin{table*}
    \centering
    \begin{tabular}{|c|c | c | c | c | c | c | c | c | c | c | c |}
    \hline
         \multirow{3}{*}{BS} & \multicolumn{9}{|c|}{Training Data Ratio} \\ 
         \cline{2-10}
          &  \multicolumn{3}{|c|}{0.01} & \multicolumn{3}{|c|}{0.001} & \multicolumn{3}{|c|}{0.0001}\\ \cline{2-10}
         & Baseline & RM Train & RM Fixed & Baseline & RM Train & RM Fixed & Baseline & RM Train & RM Fixed \\ \hline
         128 & \multirow{4}{*}{0.812} & \textbf{0.817} & 0.714 & \multirow{4}{*}{0.6447} & \textbf{0.7207} & 0.7059 & \multirow{4}{*}{0.4871} & 0.5765 & \textbf{0.662} \\ 
        256 & & \textbf{0.8172} & 0.6986 & & \textbf{0.7165} & 0.6903 & & 0.5841 & \textbf{0.6532}\\
        512 & & \textbf{0.8203} & 0.6973 & & \textbf{0.7085} & 0.6923 & & 0.5587 & \textbf{0.6504} \\ 
        1024 & & \textbf{0.8195} & 0.6807 & & \textbf{0.7068} & 0.6713 & & 0.571 & \textbf{0.6304} \\ \hline
    \end{tabular}
    \caption{Best validation accuracy for favorite genre prediction on MovieLens-20M. Segment masking and permutation have similar performance to random masking.}
    \label{tab:results-fg-movielens20m}
\end{table*}

Table~\ref{tab:results} and Table~\ref{tab:results-RM-ratios} present the best validation accuracy for the sequence classification tasks on MovieLens 1M, trained with 1\% of the training data. The item embedding dimension is 16 and convolution filters sizes are [32, 32]. The Barlow Twins-based model consistently outperforms the baselines (dual encoder with fixed/trainable weights and a model trained from scratch) across all tasks. Notably, using fixed Barlow Twins weights generally achieves higher accuracy than fine-tuning the weights, suggesting that the pre-trained representations are already highly informative and less prone to overfitting on extremely limited labeled training data (see more details in Figure~\ref{fig:plot}).


For favorite genre and occupation prediction, the improvement from using Barlow Twins is substantial. In the age prediction task, while the advantage is less pronounced, Barlow Twins-based models still generally outperform the baselines. 

Interestingly, models initialized with dual encoder representations (DE train, DE fixed) sometimes underperform even the model trained from scratch (Baseline), suggesting poor transferability of the dual encoder representation to different tasks.



Tables \ref{tab:results-fg-movielens20m} and \ref{tab:results-fc-yelp} report favorite category prediction results for MovieLens 20M and Yelp, respectively. We focus on random masking with p=0.2, which consistently yielded the best performance across all classification tasks on MovieLens 1M. Similar to MovieLens 1M, using Barlow Twins weights is superior to training from scratch. Furthermore, as the proportion of training data decreases, the advantage of using fixed weights becomes more pronounced.


\subsection{Next-item Prediction}

\begin{figure*}
    \centering
    \includegraphics[width=0.225\linewidth]{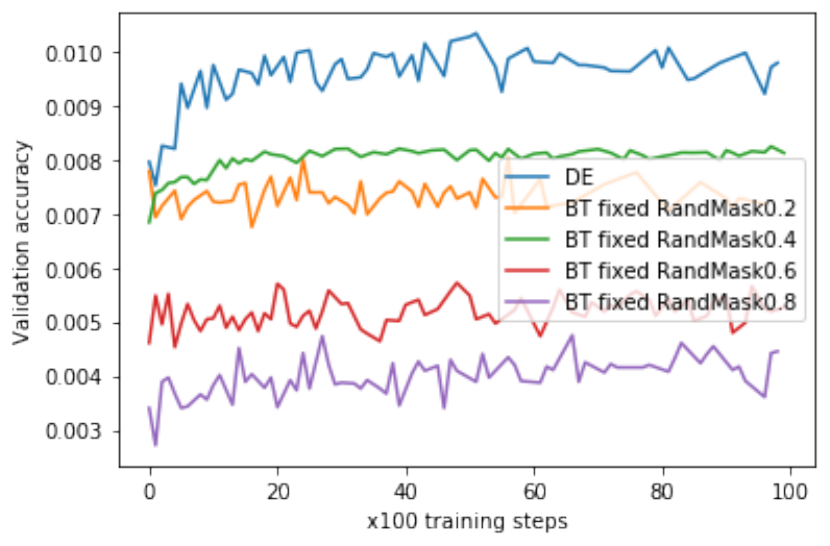}
    \includegraphics[width=0.225\linewidth]{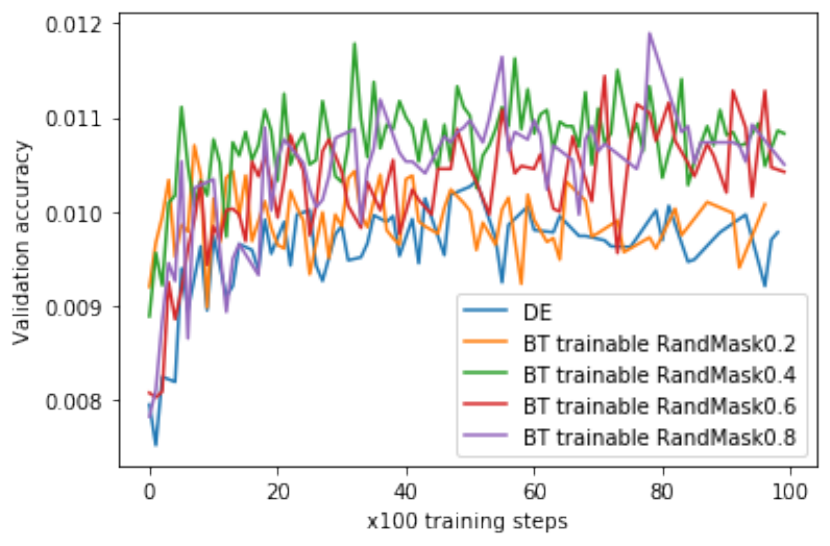}
    \includegraphics[width=0.225\linewidth]{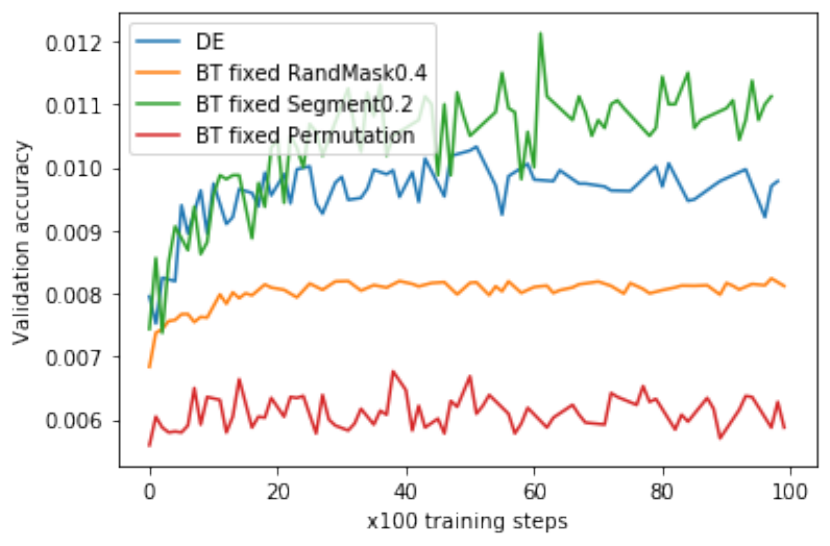}
    \includegraphics[width=0.225\linewidth]{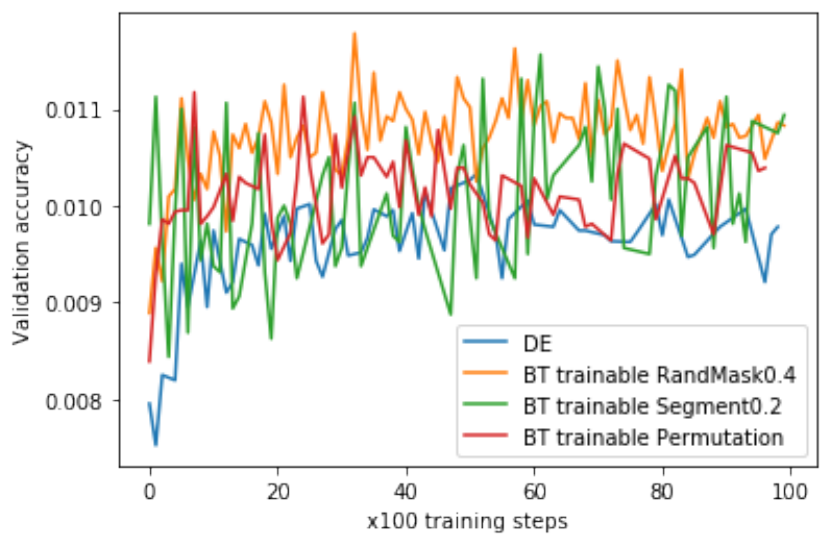}
    \includegraphics[width=0.225\linewidth]{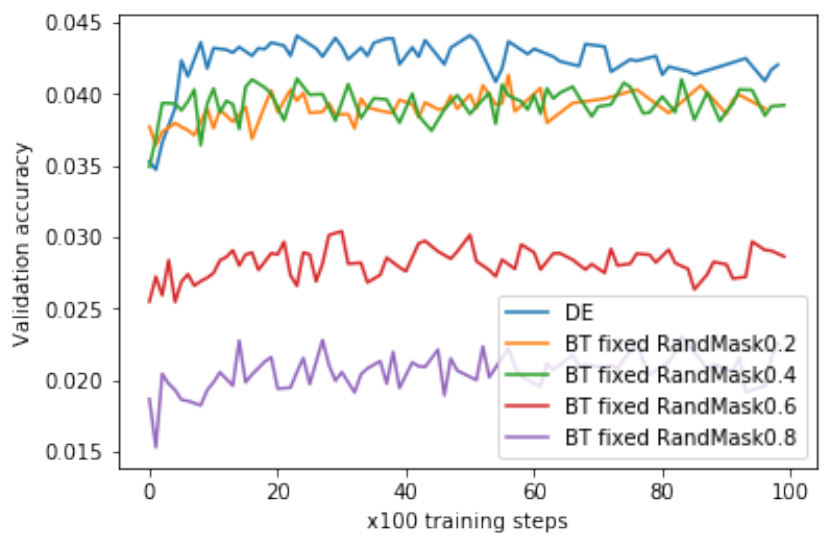}
    \includegraphics[width=0.225\linewidth]{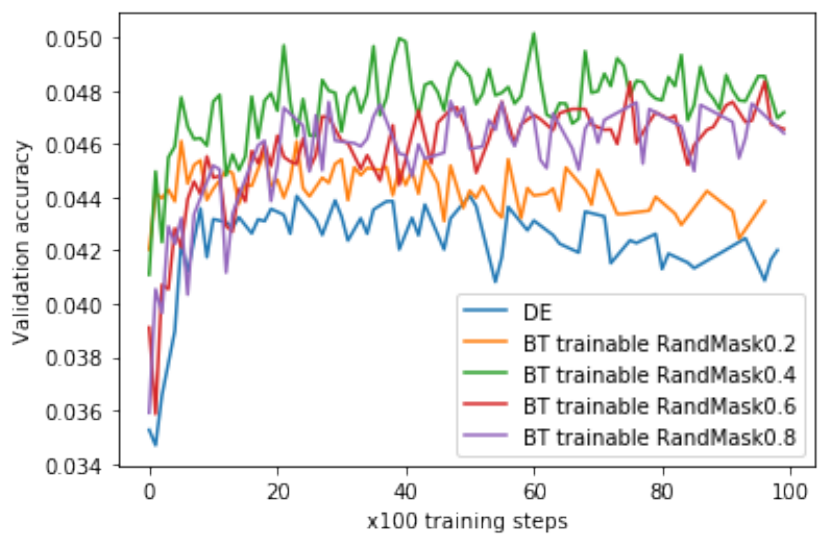}
    \includegraphics[width=0.225\linewidth]{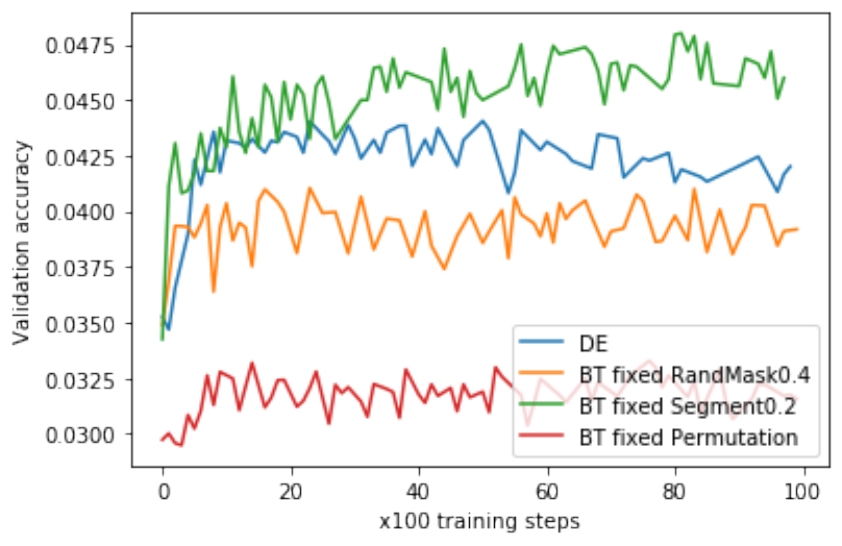}
    \includegraphics[width=0.225\linewidth]{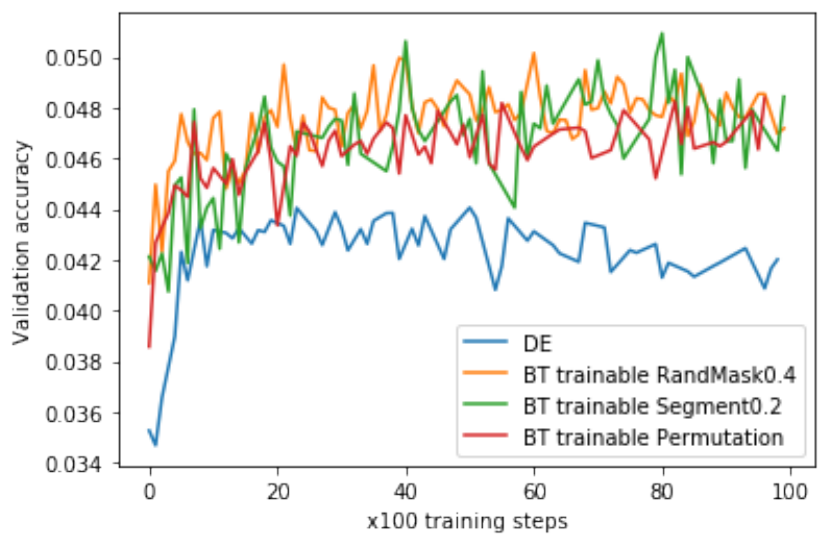}
    \includegraphics[width=0.225\linewidth]{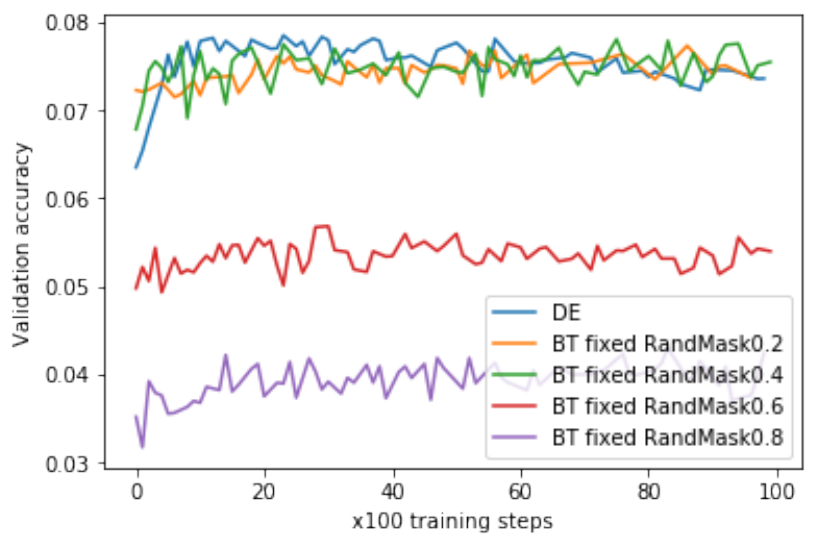}
    \includegraphics[width=0.225\linewidth]{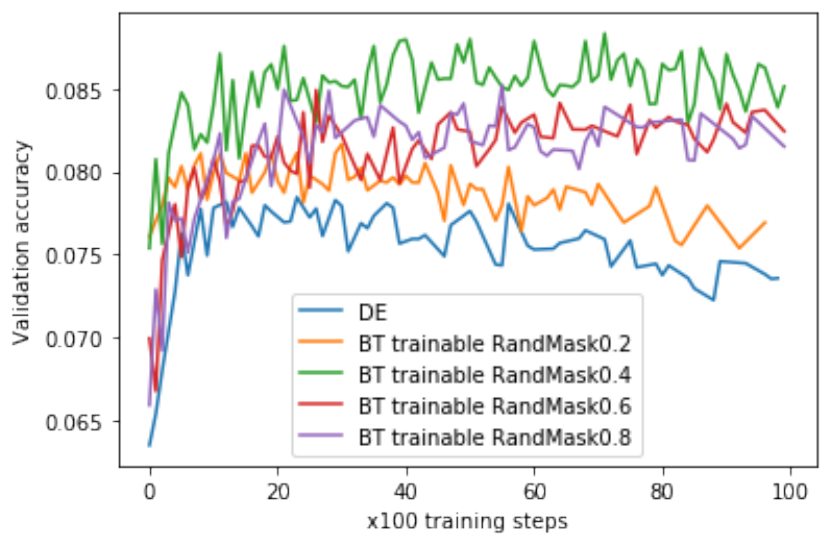}
    \includegraphics[width=0.225\linewidth]{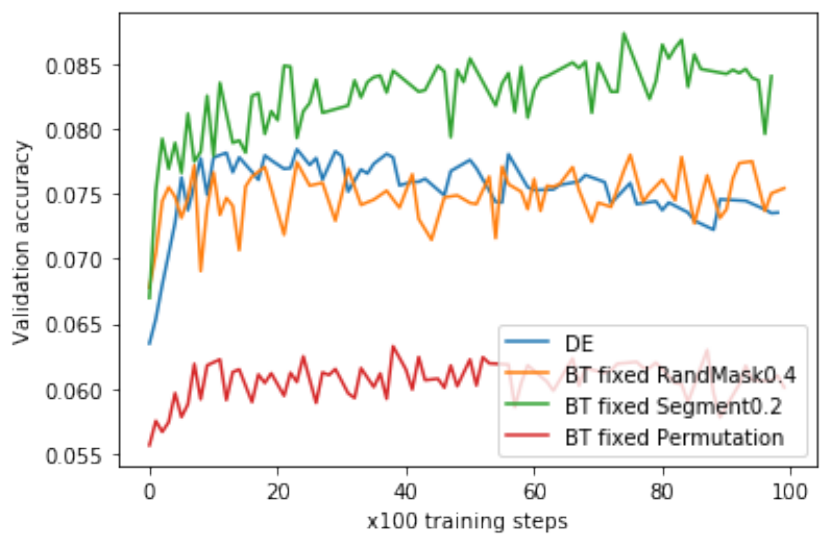}
    \includegraphics[width=0.225\linewidth]{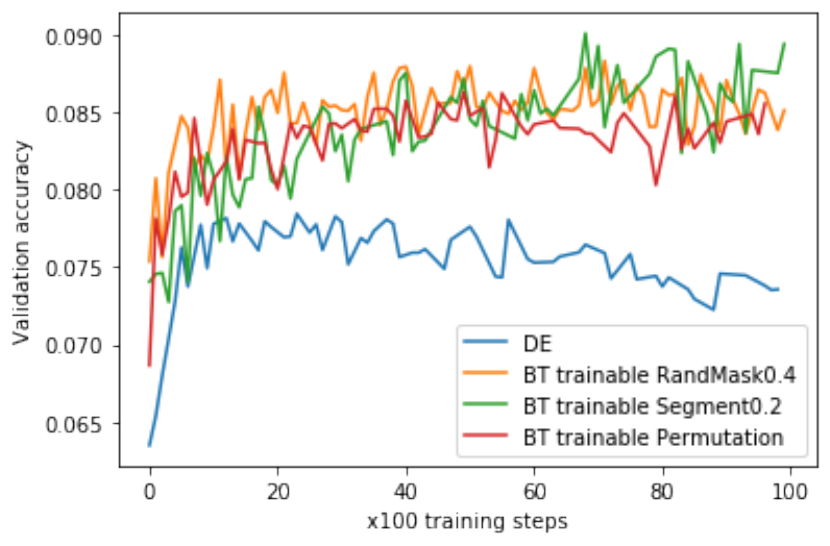}
    \caption{Validation recall (hit-ratio) for next movie prediction on MovieLens-1M. Barlow Twins/dual-encoder batch size=128. Three types of augmentations for Barlow Twins. \textit{Top to bottom}: top 1, 5, 10 recall. \textit{Left to Right}: random masking+ fixed weight, random masking + trainable weight, all augmentations + fixed weight, all augmentations + trainable weight.}
    \label{fig:next-movie-1M}
\end{figure*}

To assess the effectiveness of Barlow Twins representations on the next-item prediction task, we initialize a dual-encoder model with pretrained Barlow Twins weights and compare it to a dual encoder model trained solely for this task. The evaluation metric is top-$k$ recall (hit-ratio) with $k\in \{1,5,10\}$, which measures the percentage of cases where the ground truth next item appears within the top $k$ recommendations based on cosine similarity.


For MovieLens 1M, Figure~\ref{fig:next-movie-1M} presents the validation curves for the next-item prediction task using different augmentation strategies for Barlow Twins, with a batch size of 128. Remarkably, even with fixed Barlow Twins weights (i.e., only training a small MLP head in the context tower), the model with segment masking at p=0.2 surpasses the performance of the dual encoder baseline that was trained specifically for this task. Further improvements are achieved by fine-tuning the Barlow Twins weights.

Table \ref{tab:results-next-item-movielens20m-yelp} reports the top-5 and top-10 recalls for MovieLens 20M and Yelp. Due to the large item space in these datasets (see Table \ref{tab:dataset}), top-1 recall becomes extremely challenging and is therefore omitted. While we no longer observe the fixed-weight Barlow Twins model outperforming the baseline, models initialized with Barlow Twins and then fine-tuned still achieve significant improvements over the dual encoder baseline.


\begin{table}
    \centering
    \begin{tabular}{|c |c | c | c | c | c | c | c | c |}
    \hline
         \multirow{3}{*}{BS} & \multicolumn{6}{|c|}{Training Data Ratio} \\ 
         \cline{2-7}
          &  \multicolumn{3}{|c|}{0.01} & \multicolumn{3}{|c|}{0.05}\\ \cline{2-7}
         & Baseline & RM T & RM F & Baseline & RM T & RM F\\ \hline
         128 & \multirow{3}{*}{0.2082} & 0.208 & \textbf{0.2113} & \multirow{3}{*}{0.4061} & \textbf{0.4468} & {0.2193}\\ 
        256 & & 0.2081 & \textbf{0.2177} & & \textbf{0.4536} & 0.2237 \\
        512 & & 0.217 & \textbf{0.2142} & & \textbf{0.4438} & 0.2241 \\ \hline
    \end{tabular}
    \caption{Best validation accuracy for favorite category prediction on Yelp dataset. RM T: random masking trainable, RM F: random masking fixed. Segment masking and permutation have similar performance to random masking. }
    \label{tab:results-fc-yelp}
\end{table}

\begin{table*}
    \centering
    \begin{tabular}{|c |c | c | c | c |c | c | c | c | c |}
    \hline
         Dataset & Metric & SSL BS & RM Train & RM Fixed & SM Train & SM Fixed & Per Train & Per Fixed & DE (baseline)\\ \hline
         \multirow{8}{*}{MovieLens 20M } &\multirow{4}{*}{Top-5} & 128 & 0.0291 & 0.0183 &\textbf{ 0.0301} & 0.0218 & 0.0291&0.0183 & \multirow{4}{*}{0.0265} \\
         & & 256 &0.0264 &0.0148 & \textbf{0.0301 }& {0.0232} &0.0264 & 0.0148 &\\
         & & 512 & 0.0292 & 0.0134 & \textbf{0.0303} & {0.0191} & 0.0292 &0.0134 &\\
         & & 1024 & {0.0294} & {0.0132} & \textbf{0.0298} & 0.021 & 0.0294 & 0.0132 &\\\cline{2-10}
         &\multirow{4}{*}{Top-10} & 128 & 0.0544 & 0.0352 &\textbf{ 0.0557} & 0.0404 & 0.0544& 0.0352 & \multirow{4}{*}{0.0503} \\
         & & 256 & 0.0505 & 0.0276 & \textbf{0.0562 }& {0.0432} & 0.0505 & 0.0276 &\\
         & & 512 & 0.0554 & 0.0268 & \textbf{0.0555} & {0.0372} & 0.0554 & 0.0268 &\\
         & & 1024 & \textbf{0.0548} & {0.0259} & \textbf{0.0548} & 0.0395 & 0.0548 & 0.0259 &\\\hline
        \multirow{8}{*}{Yelp} &\multirow{4}{*}{Top-5} & 128 & 0.0753 & 0.0483 & \textbf{0.0756} & 0.0608 & 0.0639& 0.0309 & \multirow{4}{*}{0.0578} \\
         && 256 & \textbf{0.0751 }& 0.0483& 0.0737 & {0.066} & 0.0605 & 0.0388 &\\
         && 512 & \textbf{0.0768 }& 0.0512 & {0.0751} & {0.0577} & 0.068 &0.0505 &\\
         && 1024 & 0.0709 & {0.0458} & \textbf{0.0709} & 0.0492 & 0.0619 & 0.048 &\\\cline{2-10}
          &\multirow{4}{*}{Top-10} & 128 & 0.0935 & 0.07 & \textbf{0.0954} & 0.0847 & 0.0817& 0.0492 & \multirow{4}{*}{0.0717} \\
         && 256 & \textbf{0.0931 }& 0.0669 & \textbf{0.0931} & {0.0899} & 0.0776 & 0.0601 &\\
         && 512 & \textbf{0.0966 }& 0.0672 & {0.0939} & {0.076} & 0.0867 & 0.0708 &\\
         && 1024 & 0.0889 & {0.0577} & \textbf{0.0891} & 0.0616 & 0.0795 & 0.0647 &\\\hline
    \end{tabular}
    \caption{Best validation top-5 and top-10 recall (HR) of next item prediction task on MovieLens 20M and Yelp datasets. SSL BS: SSL batch size, RM: random masking, SM: segment masking, Per: permutation, DE: dual encoder, Train: trainable. The highest accuracy in each row is in bold.}
    \label{tab:results-next-item-movielens20m-yelp}
\end{table*}

\section{Discussion}
\subsection{Effect of Different Augmentation Methods}
We focus our discussion on the results obtained with \textit{fixed weights} in the downstream tasks, as this directly reflects the quality of the learned representations.

For random masking, a high masking ratio ($p=0.6$ or $0.8$)consistently leads to poor performance (see the first column of Figure~\ref{fig:next-movie-1M}. We argue that when the ratio is too high, a lot of information in the sequence is discarded and thus it is hard to learn useful representations. While $p=0.2$ and $p=0.4$ achieve decent performance for sequence-level classification tasks, they lead to worse performance on next-item prediction tasks compared to the dual encoder baseline.


Segment masking with $p=0.2$ emerges as the most effective augmentation method overall. Notably, it is the only method that outperforms the dual encoder baseline in next-item prediction. Note that it outperforms random masking with the same mask ratio. This suggests that recovering a contiguous subsequence, rather than isolated items, fosters a deeper understanding of user behavior. We hypothesize that segment masking forces the model to learn more about user intentions, habits, and preferences, whereas recovering isolated masked items may rely more on local context.


Permutation, despite being suitable for position-invariant tasks like favorite genre prediction, generally does not lead to improved performance. This observation suggests that maintaining the temporal order of actions in user sequences is crucial for capturing meaningful patterns and understanding user behavior. Disrupting this temporal order may hinder the model's ability to learn relevant representations.


\subsection{Effect of SSL on Downstream Tasks Training}

\begin{figure*}[ht]
    \centering
    \includegraphics[width=0.35\linewidth]{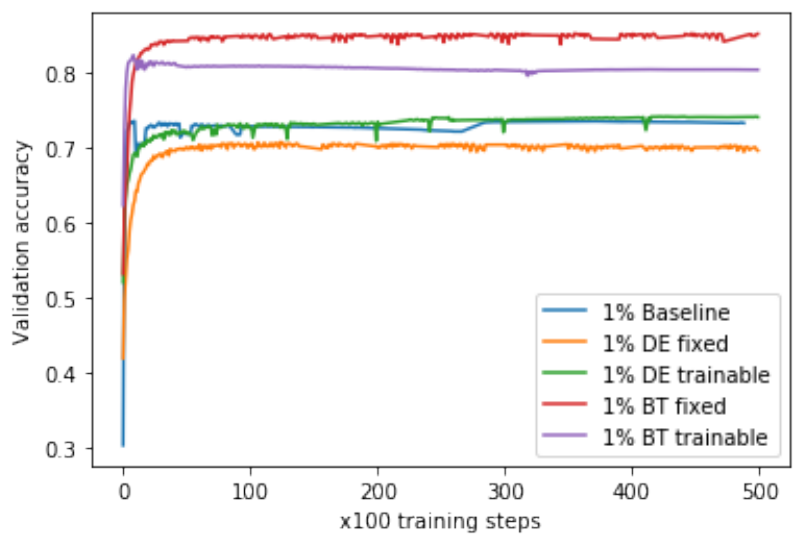} 
    \includegraphics[width=0.35\linewidth]{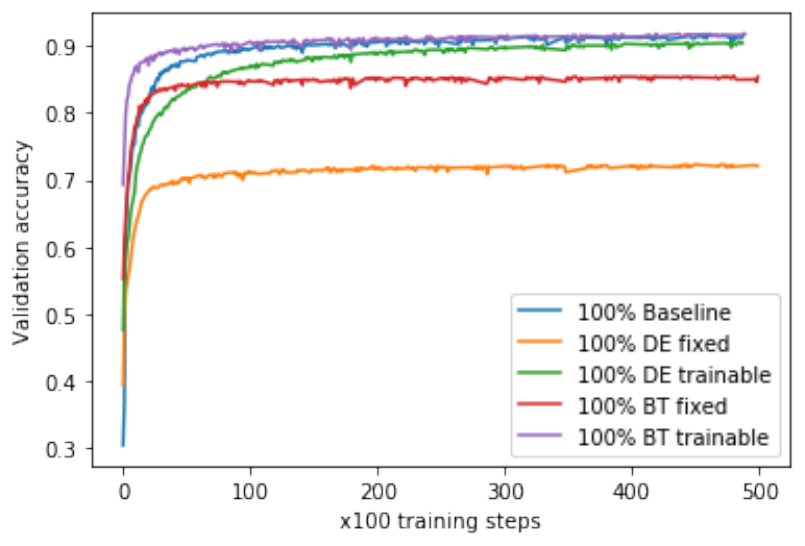}
    \caption{Favorite genre prediction with 1\% (left) and 100\% (right) training data. The batch size for SSL pretraining is 1024.}
    \label{fig:plot}
\end{figure*}

Figure \ref{fig:plot} illustrates the validation curves for favorite genre prediction on MovieLens 1M with 1\% and 100\% of the training data. With only 1\% of the data (less than 8k sequences), while the validation set is over 10 times larger, models with trainable weights suffer from overfitting. This is further confirmed in Table \ref{tab:results-final}, which shows that using fixed weights from Barlow Twins consistently achieves the best final validation accuracy. Notably, the performance drop from the best validation accuracy to the final accuracy is modest when using fixed weights (comparing Table~\ref{tab:results} and Table~\ref{tab:results-final}), highlighting the stability of this approach.

These results underscore the effectiveness of our Barlow Twins-based pre-training in learning robust representations that generalize well to downstream tasks, even in scenarios with extremely limited labeled data. By leveraging the knowledge learned from unlabeled data, we can effectively mitigate overfitting and achieve superior performance compared to training models from scratch or fine-tuning all layers. This finding highlights the potential of our approach for real-world applications where labeled data is scarce.

\begin{table*}
    \centering
    \begin{tabular}{|c | c | c | c |c | c | c | c |}
    \hline
         Task  & SSL batch size & BT trainable & BT fixed & DE trainable & DE fixed & Baseline\\ \hline
         \multirow{4}{*}{FG } & 128 & 0.7885 & \textbf{0.8394} & 0.7314 & 0.7122 & \multirow{4}{*}{0.7223} \\
         & 256 & 0.7936 & \textbf{0.8451} & 0.7461 & 0.7041 &\\
         & 512 & 0.7894 &\textbf{ 0.8385 }& 0.7528 & 0.7027 &\\
         & 1024 & 0.8028 & \textbf{0.8494}  & 0.7402 & 0.6971 &\\\hline
         \multirow{4}{*}{Occ} & 128 & 0.1317 & \textbf{0.1522} & 0.1258 & 0.1296 & \multirow{4}{*}{0.1293} \\
         & 256 & 0.1308 & \textbf{0.1535} & 0.1266 & 0.1356 &\\
         & 512 & 0.1313 &\textbf{0.1523}& 0.1199 & 0.1275 &\\
         & 1024 & 0.1307 & \textbf{0.1515} & 0.1282 & 0.1298 &\\\hline
    \end{tabular}
    \caption{Final validation accuracy of sequence-level classification for different models with 1\% training data on MovieLens-1M.}
    \label{tab:results-final}
\end{table*}

With 100\% of the training data, using fixed weights leads to suboptimal performance, as the model cannot adapt to the specific downstream task. In this scenario, the final validation accuracy of both the Barlow Twins-initialized model and the baseline trained from scratch gradually converge, as expected when sufficient labeled data is available. However, we observe a key distinction: the Barlow Twins-initialized model achieves this convergence much faster than the baseline. Additionally, it consistently outperforms the model initialized with pre-trained dual encoder weights, which even underperforms the model trained from scratch. This finding further demonstrates the superior quality and transferability of representations learned through Barlow Twins pre-training.


In next-item prediction, Barlow Twins pre-training also mitigates overfitting. This is most evident in the last column of Figure \ref{fig:next-movie-1M}, where the accuracy of the dual encoder baseline gradually declines over epochs, while the performance of Barlow Twins pretrained models remains stable or even slightly improves.

\subsection{Effect of SSL Batch Size} Our results (Tables~\ref{tab:results} to~\ref{tab:results-final}) indicate that small batch sizes do not significantly impair performance on either sequence-level classification or next-item prediction tasks. In fact, smaller batch sizes occasionally lead to higher performance (e.g. Table~\ref{tab:results} for favorite genre prediction with fixed-weight segment masking; Table~\ref{tab:results-next-item-movielens20m-yelp} for Yelp next-item prediction with trainable segment masking). This observation aligns with the findings in the original Barlow Twins paper~\cite{ZbontarJMLD21}. Importantly, smaller batch sizes offer practical advantages in reducing computational resource requirements and accelerating model convergence.

\subsection{Item Embedding Visualization}
\begin{figure*}[t]
    \centering
    \includegraphics[width=0.3\linewidth]{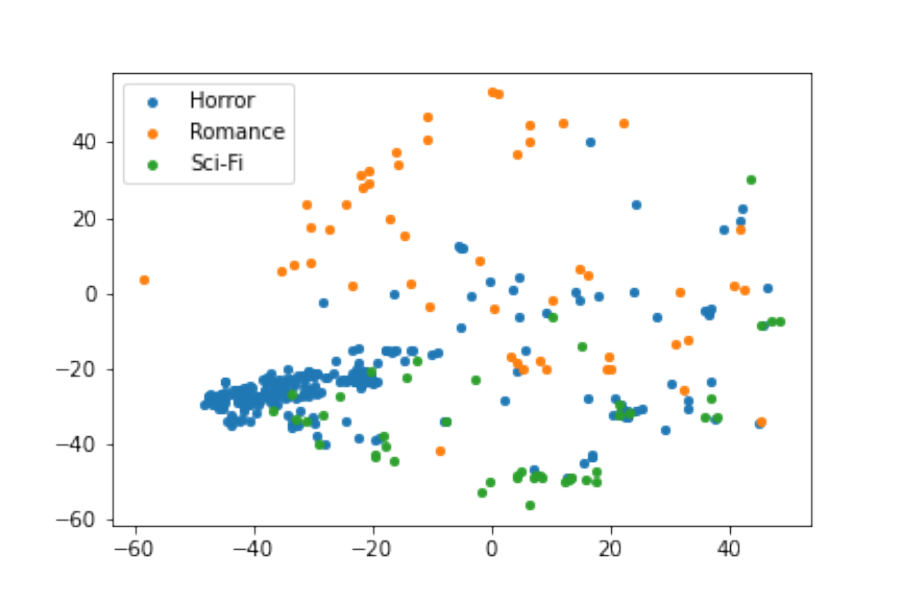}
    \includegraphics[width=0.3\linewidth]{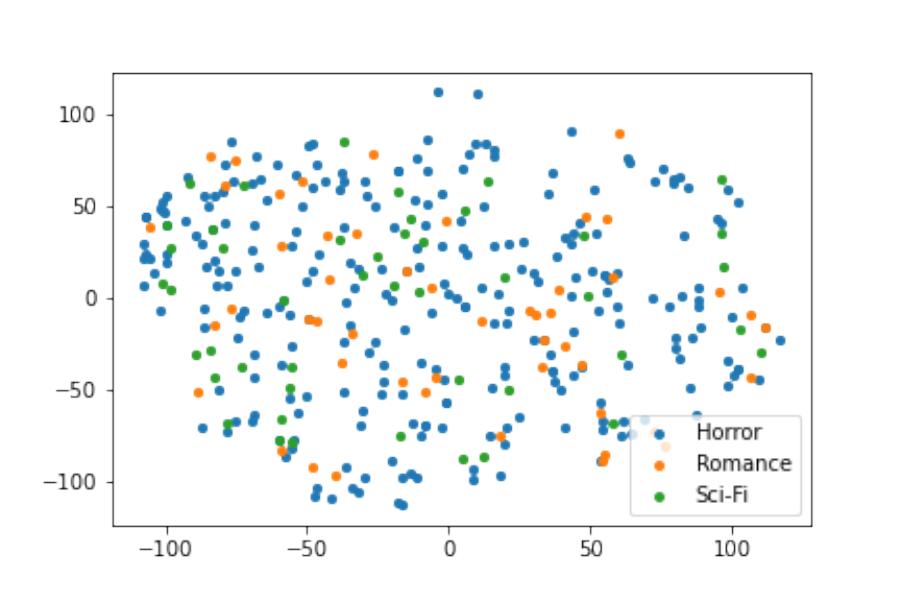}
    \includegraphics[width=0.3\linewidth]{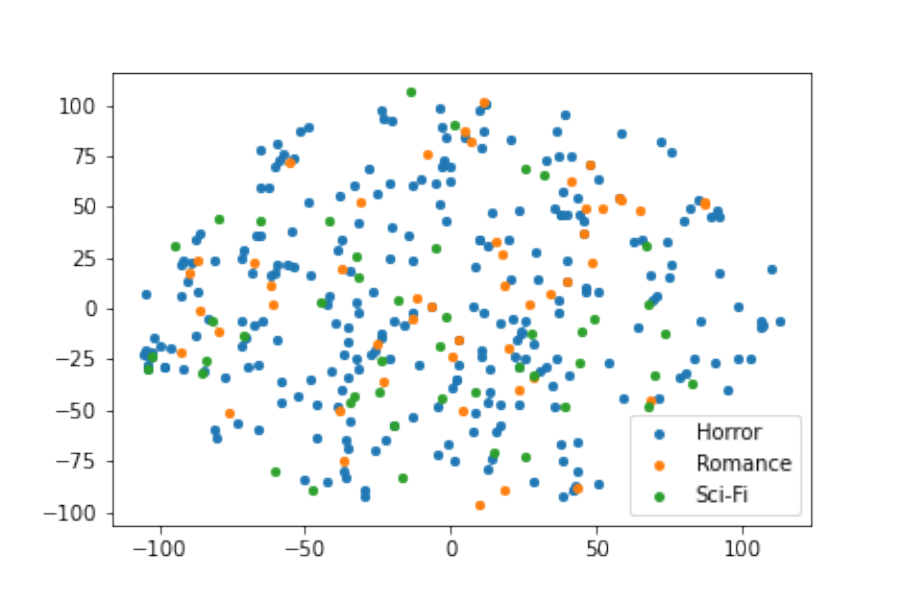}
    \caption{t-SNE plots of movie embeddings from 3 movie genres. Left: dual encoder. Middle: Barlow Twins with random masking. Right: Barlow Twins with segment masking.}
    \label{fig:tSNE}
\end{figure*}

To qualitatively evaluate the learned item embeddings, we visualize the t-SNE plots (Figure~\ref{fig:tSNE}) of the movie embeddings from three distinct genres (romance, horror, sci-fi) obtained from dual encoder and Barlow Twins (with $p=0.2$ random masking and segment masking) trained on MovieLens 1M dataset. Intuitively, these genres should form distinct clusters in a well-learned item embedding space. 

While the dual encoder model effectively separates the three genres, the Barlow Twins embeddings exhibit less distinct clustering. This observation suggests that while Barlow Twins excels at learning high-level sequence representations, it may not be as effective at optimizing item-level embeddings compared to the dual encoder, which directly supervises the item tower during contrastive learning.This disparity may stem from the Barlow Twins loss being applied only at the end of the model, resulting in weaker backpropagation signals for the initial item embedding layers.



This observation suggests that further improvements in sequence-level representations may be achievable by enhancing the quality of the item embeddings. Potential strategies for this include incorporating reconstruction tasks for masked actions, similar to BERT~\cite{DevlinCLT19}, or jointly training Barlow Twins with a next-item prediction objective, as done in prior works~\cite{xie2022contrastive,ChenLLMX22,XiaHHLYK23}.

\section{Conclusion}

In this work, we have explored the application of Barlow Twins-based self-supervised learning to learn general-purpose sequence-level representations for user modeling tasks. Our experiments demonstrate that adapting Barlow Twins to user sequence data yields several practical benefits. First, Barlow Twins learns versatile sequence-level representations that effectively transfer to various downstream tasks. Second, our approach mitigates overfitting when fine-tuning on limited labeled data, leading to more stable and accurate downstream models. Third, even with abundant labeled data, Barlow Twins pre-training accelerates convergence and can improve final performance on downstream tasks.

While our results highlight the potential of Barlow Twins for user sequence modeling, a limitation of our current approach is its focus on sequence-level rather than item-level representations. Future work could investigate techniques for jointly optimizing both levels of representation, potentially by incorporating reconstruction tasks for masked items or integrating a next-item prediction objective into the Barlow Twins framework. This could further enhance the applicability and effectiveness of Barlow Twins-based SSL for personalized recommendation systems and other user modeling tasks.

\begin{acks}
\end{acks}

\bibliographystyle{WWW2024/ACM-Reference-Format}
\bibliography{references}

\appendix


\end{document}